\documentclass[9pt]{elife}

\usepackage{lipsum} 
\usepackage[version=4]{mhchem}
\usepackage{siunitx}
\usepackage{mathrsfs}
\DeclareSIUnit\Molar{M}

\usepackage{nicefrac}
\usepackage{mathtools}

\title{Interkinetic nuclear migration in the zebrafish retina as a diffusive process}

\author[1\authfn{1}]{Afnan Azizi}
\author[2\authfn{1}]{Anne Herrmann}
\author[3]{Yinan Wan}
\author[1]{Salvador J. R. P. Buse}
\author[3]{Philipp J. Keller}
\author[2*]{Raymond E. Goldstein}
\author[1*]{William A. Harris}
\affil[1]{Department of Physiology, Development and Neuroscience, University of Cambridge, 
Cambridge CB2 3DY, United Kingdom}
\affil[2]{Department of Applied Mathematics and Theoretical Physics, Centre for 
Mathematical Sciences, University of Cambridge, Cambridge CB3 0WA, United Kingdom}
\affil[3]{Howard Hughes Medical Institute, Janelia Research Campus, Ashburn, VA 20147, USA}

\corr{wah20@cam.ac.uk}{WAH}
\corr{r.e.goldstein@damtp.cam.ac.uk}{REG}

\contrib[\authfn{1}]{These authors contributed equally to this work}

\begin{document}

\maketitle

\begin{abstract}
A major hallmark of neural development is the oscillatory movement of nuclei between the 
apical and basal surfaces of the neuroepithelium during the process of interkinetic nuclear 
migration (IKNM). Here, we employ long-term, rapid lightsheet and two-photon imaging of 
zebrafish retinas \textit{in vivo} during early development to uncover the physical 
processes that govern the behavior of nuclei during IKNM. These images allow the capture 
of reliable tracks of nuclear movements and division during early retinogenesis for many 
tightly packed nuclei. These tracks are then used to create and test a theory of retinal 
IKNM as a diffusive process across a nuclear concentration gradient generated by the 
addition of new nuclei at the apical surface. The analysis reveals the role of nuclear 
packing at the apical surface on the migration dynamics of nuclei, provides a robust 
quantitative explanation for the distribution of nuclei across the retina, and may have 
implications for stochastic fate choice in this system.
\end{abstract}

\section{Introduction}

The vertebrate nervous system arises from a pseudostratified epithelium within which elongated
proliferating cells contact both the apical and basal surfaces. Within these cells, striking 
nuclear movements take place during the proliferative phase of neural development. More than 80 
years ago, these movements, termed interkinetic nuclear migration (IKNM), were shown to occur in 
synchrony with their cell cycle \citep{Sauer_mitosis_1935}. Under normal conditions, nuclei of
proliferating cells undergo mitosis (M) exclusively at the apical surface. During the first gap 
phase (G1) of the cell cycle, nuclei migrate away from this surface to reach more basal positions 
by S-phase, when DNA is replicated. In the second gap phase (G2), nuclei migrate rapidly toward the 
apical surface where they divide again \citep{Del_Bene_Interkinetic_2011,Sauer_mitosis_1935,Baye_Interkinetic_2007,Leung_Apical_2011,Kosodo_Regulation_2011,Norden_Actomyosin_2009}. 
The molecular mechanisms that drive the rapid nuclear movement in G2 have been investigated in 
a number of tissues \citep{Norden_Pseudostratified_2017}. For instance, in the mammalian 
cortex they are thought to involve microtubules as well as various microtubule motors and 
actomyosin \citep{Xie_Cep120_2007,Tsai_Dual_2007}. In the zebrafish retina, it appears to be 
the actomyosin complex alone that moves the nuclei to the apical surface during G2 \citep{Norden_Actomyosin_2009,Leung_Apical_2011}. In contrast, the nuclear movements during the 
majority of the cell cycle, in G1 and S phases, have been less thoroughly examined. Although 
similar molecular motors have been implicated \citep{Schenk_Myosin_2009,Tsai_Kinesin_2010}, 
the underlying processes remain elusive.

Importantly, IKNM is known to affect morphogenesis and cell differentiation in neural tissues
\citep{Spear_Interkinetic_2012}, as retinas with perturbed IKNM are known to develop prematurely 
and to display abnormalities in cell composition \citep{Del_Bene_Regulation_2008}. Given this 
regulatory involvement of IKNM in retinal cell differentiation, a deeper understanding of the 
nuclear movements remains a major prerequisite for insights into the development of neural systems. 
On a phenomenological level, the movements of nuclei during the G1 and S phases have been shown 
to resemble a stochastic process in the zebrafish retina \citep{Norden_Actomyosin_2009,Leung_Apical_2011} . During 
these periods, individual nuclei switch between apical and basal movements at random intervals, 
leading to considerable variability in the maximum basal position they reach during each cell 
cycle \citep{Baye_Interkinetic_2007}. Similarly, in the mammalian cerebral cortex, the considerable 
internuclear variability in IKNM leads to nuclear positions scattered throughout the entire
neuroepithelium in S-phase  \citep{Sidman_Cell_1959,Kosodo_Regulation_2011}. The high variability 
in the movements of nuclei during G1 and S phases of the cell cycle suggests that passive, rather 
than active, molecular processes are a main driver of basal migration. This hypothesis was supported 
by experiments demonstrating similarly variable basalward-biased migration of nuclear-sized 
microbeads inserted in between cells during IKNM in the mouse cortex \citep{Kosodo_Regulation_2011}.
Various possible explanations for these passive processes have been put forward. These suggestions 
include the possibility of direct energy transfer from rapidly moving G2 nuclei
\citep{Norden_Actomyosin_2009}, as well as nuclear movements caused by apical crowding
\citep{Kosodo_Regulation_2011,Okamoto_TAG_2013}. Here, we present experiments to test these hypotheses. 

Our work relies on the tracks of closely packed nuclei of zebrafish retinal progenitor cells. The 
retina of the  oviparous zebrafish is easily accessible to light microscopy throughout embryonic
development \citep{Avanesov_Analysis_2010} and has been used for several studies of the movements 
of nuclei during IKNM 
\citep{Baye_Interkinetic_2007,Del_Bene_Regulation_2008,Norden_Actomyosin_2009,Sugiyama_Illuminating_2009,Leung_Apical_2011}. 
We find evidence for IKNM being driven by apical crowding and therefore further develop this idea 
into a mathematical model. Given the seemingly stochastic nature of individual nuclear trajectories, 
we base the model on a comparison between IKNM and a simple diffusion process. The model reveals 
the remarkable and largely overlooked importance of simple physical constraints imposed by the 
overall tissue architecture and allows us to describe accurately the global distribution of nuclei 
as a function of time within the retinal tissue. In this way, we describe IKNM as 
a tissue-wide rather than a single-cell phenomenon. In the future, this description might shed 
light on other aspects of progenitor cell biology, such as cell cycle exit and fate.

\section{Results}

\subsection{Generating image sets with high temporal resolution}

We imaged fluorescently-labeled nuclei of whole retinas of developing zebrafish at 2 min 
intervals, an optimal time period given the difficulty to track nuclei accurately over long 
times and the increased photobleaching with shorter intervals.  
We compared movies of retinas imaged at 2 min and at 20 s intervals over a period of 2 hours 
and found that the improvement in temporal resolution made no difference to our analyses.  
This suggests that it is unlikely that within each 2 min interval there were important 
intervening movements that might complicate the analysis.  

\begin{figure}
\includegraphics[width=\linewidth]{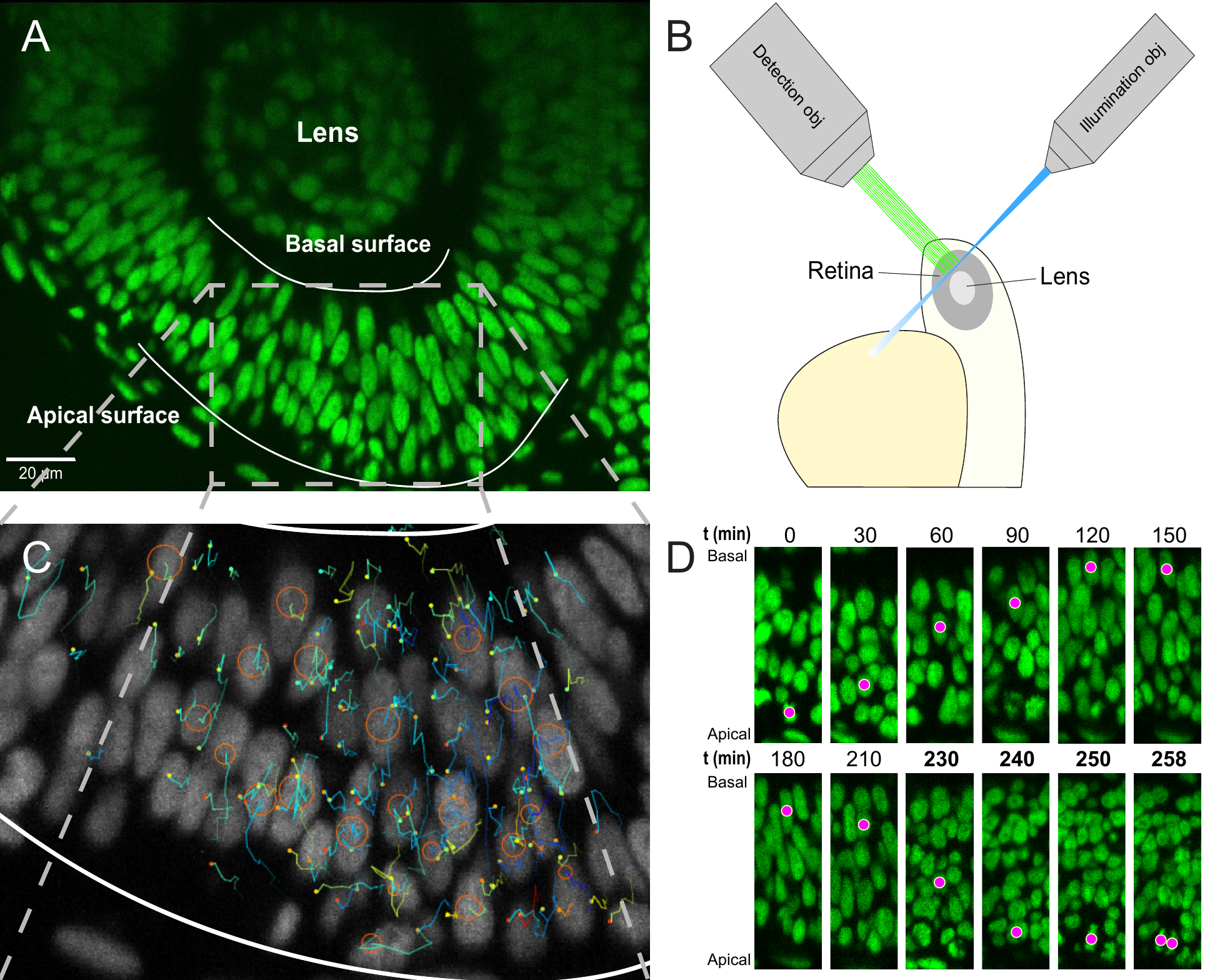}
\caption{Imaging and tracking fluorescently labeled nuclei. \textbf{(A)} A transgenic H2B-GFP 
embryonic retina imaged using lightsheet microscopy at $\sim$30 hpf. The lens, as well as apical 
and basal surfaces are indicated. \textbf{(B)} A schematic representation of single-angle 
lightsheet imaging of the retina. Laser light is focused into a sheet of light by the 
illumination objective and scans the retina. Fluorescent light is then collected by the 
perpendicular detection objective. \textbf{(C)} Track visualization and curation using the 
MaMuT plugin of Fiji. All tracks within a region of the retina are curated and visualized. 
Circles and dots represent centers of nuclei, and lines show their immediate (10 previous steps) 
track. \textbf{(D)} The position of a single nucleus within the retinal tissue from its birth to 
its eventual division. The magenta dot indicates the nucleus tracked at various time points during 
its cell cycle. The last 4 panels are at shorter time intervals to highlight the rapid movement 
of the nucleus prior to mitosis.}
\label{fig:1}
\end{figure}

To follow the nuclei of all cells within a portion of the retina we used H2B-GFP transgenic 
lines with GFP expression exclusively in the nuclei (Figure~\ref{fig:1}A). In order to achieve 
the desired temporal resolution without sacrificing image quality, fluorescence bleaching and 
sample drift must be minimized as much as possible. The retinas of H2B-GFP embryos were imaged 
using either a single-angle lightsheet microscope (see Figure~\ref{fig:1}B for a schematic) or 
an upright two-photon scanning microscope. Both of these methods yield images with minimal 
bleaching compared to other microscopic techniques \citep{svoboda_principles_2006,Stelzer_Light_2015}.
However, while the single-angle lightsheet can generate large stacks of images, it is very sensitive 
to lateral drift due to a small area of high resolution imaging. Therefore, some datasets were 
produced using two-photon microscopy, which, despite the limitations of scanning time, could 
produce areas of high resolution images of sufficient size.

Both lightsheet and two-photon microscopes produced images of at least half the retina with a 
depth of at least 50 \textmu m over several hours in 2 min intervals. The images were processed 
using a suite of algorithms \citep{Amat_Efficient_2015} to compress them to a lossless format, 
Keller Lab Block (KLB), correct global and local drift, and normalize signal intensities for 
further processing. Automated segmentation and tracking of the nuclei were carried out through 
a previously published computational pipeline that takes advantage of watershed techniques and
persistence-based clustering (PBC) agglomeration to create segments and Gaussian mixture models with 
Bayesian inference to generate tracks of nuclei through time 
\citep{Amat_Fast_2014,Amat_Efficient_2015}. Two main parameters greatly affect tracking results, 
overall background threshold and PBC agglomeration threshold. To obtain best automated tracking 
results, ground truth tracks were created for a section of the retina over 120 min and were 
compared to tracks generated over a range of these two parameters. The best combination of 
the two parameters was chosen as the one with highest tracking fidelity and lowest amount 
of oversegmentation over that interval.

The most optimal combination of parameters yielded an average linkage accuracy, from each time 
point to the next, of approximately 65\%. Hence, extensive manual curation and correction of tracks 
were required. Tracking by Gaussian mixture models (TGMM) software generates tracks that can be 
viewed and modified using the Massive Multi-view Tracker (MaMuT) plugin of the Fiji software
\citep{Wolff_Multi_2018,Schindelin_Fiji_2012}. A region of the retina with the best 
fluorescence signal was chosen and all tracks within that region were examined and any 
errors were corrected. The tracks consist of sequentially connected sets of 3D coordinates 
representing the centers of each nucleus (Figure~\ref{fig:1}C), with which their movement 
across the tissue can be mapped over time. For example, Figure~\ref{fig:1}D shows IKNM of 
a single nucleus tracked from its birth, at the apical surface of the retina, to its eventual 
division into two daughter cells.

\subsection{Analysis of nuclear tracks}
\begin{figure}
\includegraphics[width=\linewidth]{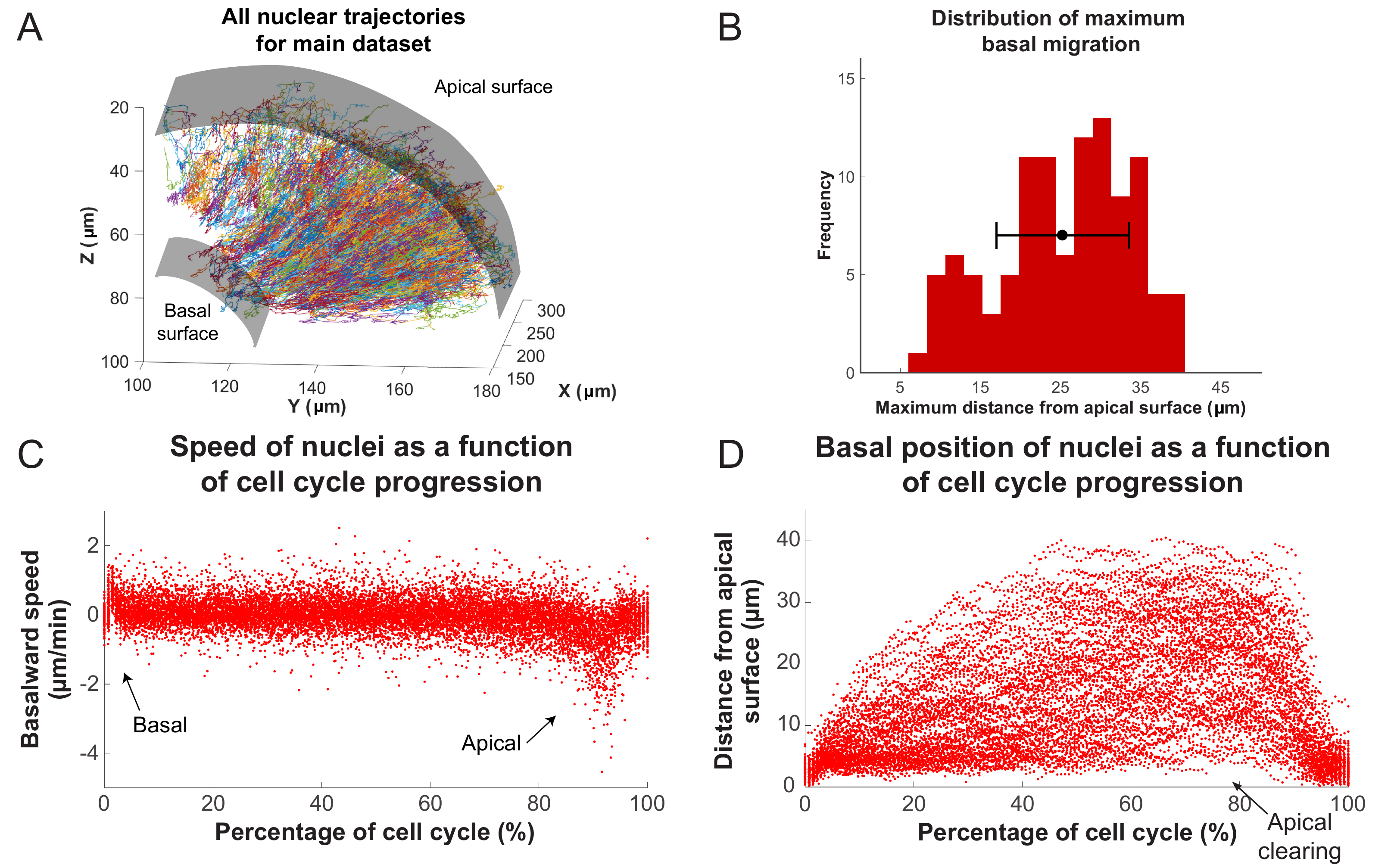}
\caption{Analysis of nuclear tracks during IKNM. \textbf{(A)} Extracted trajectories of 
nuclei in 3 dimensions. All curated tracks of the main dataset over 400 minutes in the region shown in 
Figure~\ref{fig:1}C are presented. \textbf{(B)} The distribution of maximum distances 
reached away from the apical surface by nuclei during their completed cell cycles. The 
mean and one standard deviation are shown. \textbf{(C)} The speed distribution of nuclei 
over complete cell cycles. The cell cycle lengths of all nuclei were normalized and 
superimposed to highlight the early basal burst of speed, as well as pre-division apical 
rapid migration. The speeds between these two periods are normally distributed. 
\textbf{(D)} Position of nuclei as measured by their distance from the apical surface 
over normalized cell cycle time. Even though all nuclei start and end their cell cycle 
near the apical surface, they move out across the retina to take positions in all 
available spaces, creating an apical clearing as indicated.}
\label{fig:2}
\end{figure}

This process yielded tracks for hundreds of nuclei, across various samples, over time 
intervals of at least 200 min. We used custom-written MATLAB scripts to analyze these 
tracks. The aggregated tracks of the main dataset, in Cartesian coordinates, for all tracked 
lineages is shown in Figure~\ref{fig:2}A. Single tracks for any given time interval can 
be extracted and analyzed from this collection. In order to transform the Cartesian coordinates 
of the tracks into an apicobasal coordinate system, we drew contour curves at the apical 
surface of the retina (e.g. see Figure~\ref{fig:1}A) separating RPC nuclei from the elongated 
nuclei of the pigmented epithelium. We then calculated curves of best fit (second degree 
polynomials) in both the XY and YZ planes. Assuming that the apical cortex is perpendicular 
to the apicobasal axis of each cell, displacement vectors of the nuclei at each time point 
can be separated into apicobasal and lateral components. Since, in IKNM, the apicobasal motion 
is that of interest, we used this component for our remaining analyses.
  
Figure~\ref{fig:2}C,D shows the speed and position of tracked nuclei of the same dataset, 
over the duration of their cell cycle, for all cells that went through a full cell cycle. 
While all nuclei behave similarly minutes after their birth (early G1) and before their 
division (G2), their speed of movement and displacement is highly variable for the majority 
of the time that they spend in the cell cycle (Figure~\ref{fig:2}C,D). Most daughter nuclei 
move away from the apical surface, within minutes from being born, with a clear basalward 
bias in their speed distribution (Figure~\ref{fig:2}C). This abrupt basal motion of newly 
divided nuclei has also been recently observed by others \citep{Shinoda_Elasticity_2018,Barrasso_Live_2018}. 
However, immediately after this brief period, nuclear speeds become much more equally distributed 
between basalward and apicalward, with a mean value near 0. Such a distribution is indicative 
of random, stochastic motion, which in turn leads to a large variability in the position of 
nuclei within the tissue (away from the apical surface) during the cell cycle (Figure~\ref{fig:2}B).

Interestingly, except during mitosis, we find an apical clearing of a few microns for dividing 
cells (Figure~\ref{fig:2}D). We checked to see if this was an artifact of measuring the distance 
to nuclear centers due to nuclear shape, as nuclei are rounded during M phase but are more 
elongated along the apicobasal axis at other times. We found no significant difference between 
average length of nuclear long axis when measured for nuclei right before their division compared 
to nuclei chosen randomly from any other time point within the cell cycle, indicating that this 
clearing is likely to have a biological explanation, such as the preferential occupancy of M phase 
nuclei to the apical surface during IKNM.

\subsection{Basal movement of nuclei is driven like a diffusive process}

\begin{figure}
\includegraphics[width=\linewidth]{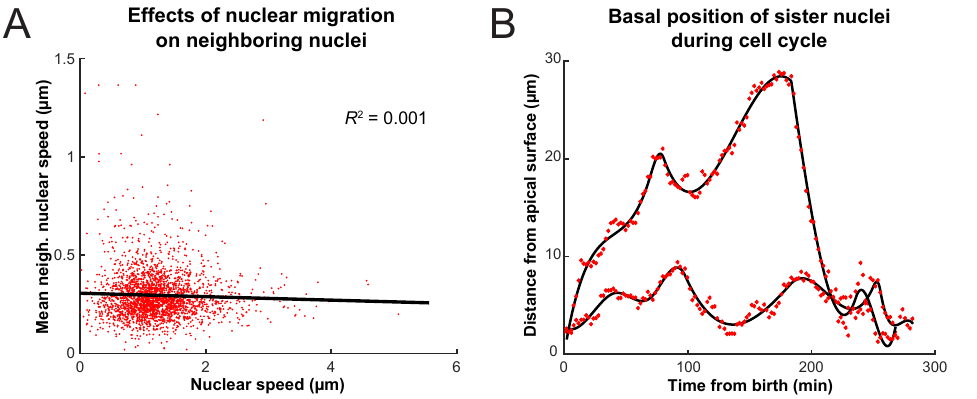}
\caption{\textbf{(A)} Average speed of nuclei neighboring a nucleus of interest as a function 
of the speed of that nucleus. \textbf{(B)} The positions of two sister nuclei at each time 
point imaged (red circles) over their complete cell cycle. The black lines are spline curves 
indicating the general trend of their movements.}
\label{fig:3}
\end{figure}

Previous work has shown that when RPCs are pharmacologically inhibited from replicating 
their DNA, their nuclei neither enter G2 nor exhibit rapid persistent apical migration 
that normally occurs during the G2 phase of the cell cycle \citep{Leung_Apical_2011,Kosodo_Regulation_2011}.  A more surprising result of these experiments 
is that the stochastic movements of nuclei in G1 and S phases also slow down considerably during 
such treatment \citep{Leung_Apical_2011}. It was, therefore, suspected that the migration of nuclei 
of cells in G2 toward the apical surface jostles those in other phases \citep{Norden_Actomyosin_2009}. 
We therefore searched our tracks for evidence of such direct kinetic interactions among nuclei 
by correlating the speed and direction of movement of single nuclei with their nearest neighbors. 
These neighbors were chosen such that their centers fell within a cylindrical volume of a height 
and base diameter twice the length of long and short axes, respectively, of an average nucleus. Figure~\ref{fig:3}A shows the lack of correlation between the speed of movement of nuclei and 
the average speed of their neighbors. We further categorized the neighboring nuclei by their 
position in relation to the nucleus of interest (along the apicobasal axis), their direction of 
movement, and whether they were moving in the same direction of the nucleus of interest or not. 
None of the resulting eight categories of neighboring nuclei showed a correlation in their 
average speed with the speed of the nucleus of interest. Furthermore, we considered the movement 
of neighboring nuclei one time point (2 min) before or one time point after the movement of the 
nucleus of interest. Yet, we still found no correlation between these time-delayed and original 
speeds. These results suggest that there does not appear to be much transfer of kinetic energy 
between neighboring nuclei.

Another hypothesis advanced for variability in basal IKNM is that 
the nuclear movements are driven by apical crowding \citep{Kosodo_Regulation_2011,Okamoto_TAG_2013}. How apical crowding might result in basal
IKNM can be understood by comparing IKNM to a diffusive process. In diffusion, a concentration 
gradient drives the average movement of particles from areas of high to areas of low 
concentration. However, despite the average movement being directed, each individual 
particle's trajectory is a random walk \citep{Reif_Fundamentals_1965}. Similarly, during 
IKNM a gradient in nuclear concentration is generated because nuclei divide exclusively 
at the apical surface. If basal IKNM were comparable to diffusion, this nuclear concentration 
gradient would be expected to result in a net movement of nuclei away from the area of high 
nuclear crowding at the apical side of the neuroepithelium 
\citep{Miyata_Interkinetic_2015, Okamoto_TAG_2013}. Indeed, in IKNM we find that each 
individual nucleus' trajectory resembles 
a random walk \citep{Norden_Actomyosin_2009}. Therefore, for the cells in the G1 and 
S phases (which account for more than 90\% of the cell cycle time in our system), IKNM has, 
at least on a phenomenological level, the main features of a diffusive process.

\begin{figure}
\includegraphics[width=\linewidth]{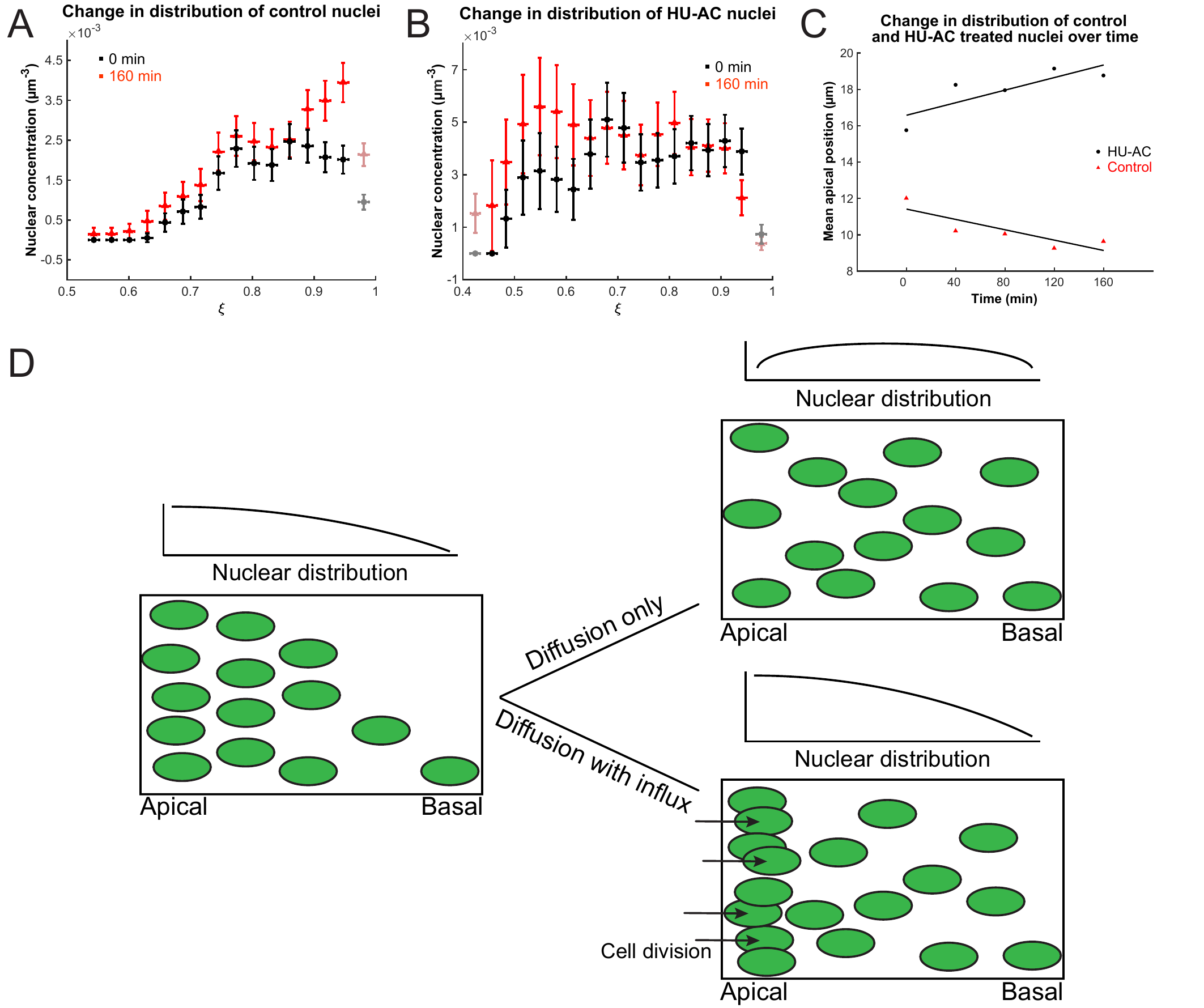}
\caption{Nuclear concentration gradient across the apicobasal axis of the retina. The 
concentration of nuclei is higher near the apical surface compared to the basal surface. 
\textbf{(A)} In the control retina the nuclear concentration gradient builds up over time. 
\textbf{(B)} Blocking apical migration and division of nuclei, by inhibiting S phase progression, 
leads to a shift in the distribution of nuclei towards the basal surface in the HU-AC treated 
retina. \textbf{(C)} The shift in the distribution of nuclei under HU-AC treatment when compared 
to the untreated retina. The number of nuclei away from the apical surface increases consistently 
over time in the absence of cell division, but remains the same when new nuclei are constantly 
added at the apical surface. \textbf{(D)} A schematic of how a diffusion model would work in 
the context of IKNM in the retina. A concentration gradient of nuclei (left) would drive the 
net movement of nuclei from the apical surface to the basal surface. However, without 
maintenance of the gradient, the drive for this net migration is lost (top right). In 
the retina, the gradient is maintained through cell divisions at the apical surface, 
modeled as a one way influx across the apical surface (bottom right), continuously 
driving the net movement basally.}
\label{fig:4}
\end{figure}

To test further whether we can indeed describe IKNM using a model of diffusion, we first asked 
what would happen to the concentration gradient if we blocked the cell cycle in S phase, 
which inhibits both the apical movement of the nuclei in G2 and mitosis at the apical surface.  
If the comparison to diffusion were valid, we expect the blockage to abolish the build-up and 
maintenance of the concentration gradient. We, therefore, compared the normally evolving 
distribution of nuclei in control retinas with those measured from retinas where the cell 
cycle was arrested at S-phase using a combination of hydroxyurea (HU) and aphidicolin (AC) \citep{Norden_Actomyosin_2009,Icha_independent_2016}. We counted the number of nuclei in a 
three dimensional section of the retina containing approximately 100 nuclei, at equal time 
intervals, starting with 120 min after drug treatment. The delay ensured that almost all 
cell divisions, from nuclei that had already completed the S phase at the time of treatment, 
had taken place. As expected from the diffusion model (Figure~\ref{fig:4}D), over the course of 160 min, the mean 
of the nuclear distribution moved further towards the basal surface in treated retinas, and 
the concentration difference between the apical and basal surfaces diminished (Figure~\ref{fig:4}A,C). 
In contrast, in control retinas the mean of the nuclear distribution moved towards the apical 
surface (Figure~\ref{fig:4}B,C) as the gradient continued to build up. Hence, these results 
support the suitability of a diffusive model to describe the basal nuclear migration during IKNM. 

\subsection{An analytical diffusion model of IKNM}

\begin{figure}
\includegraphics[width=\linewidth]{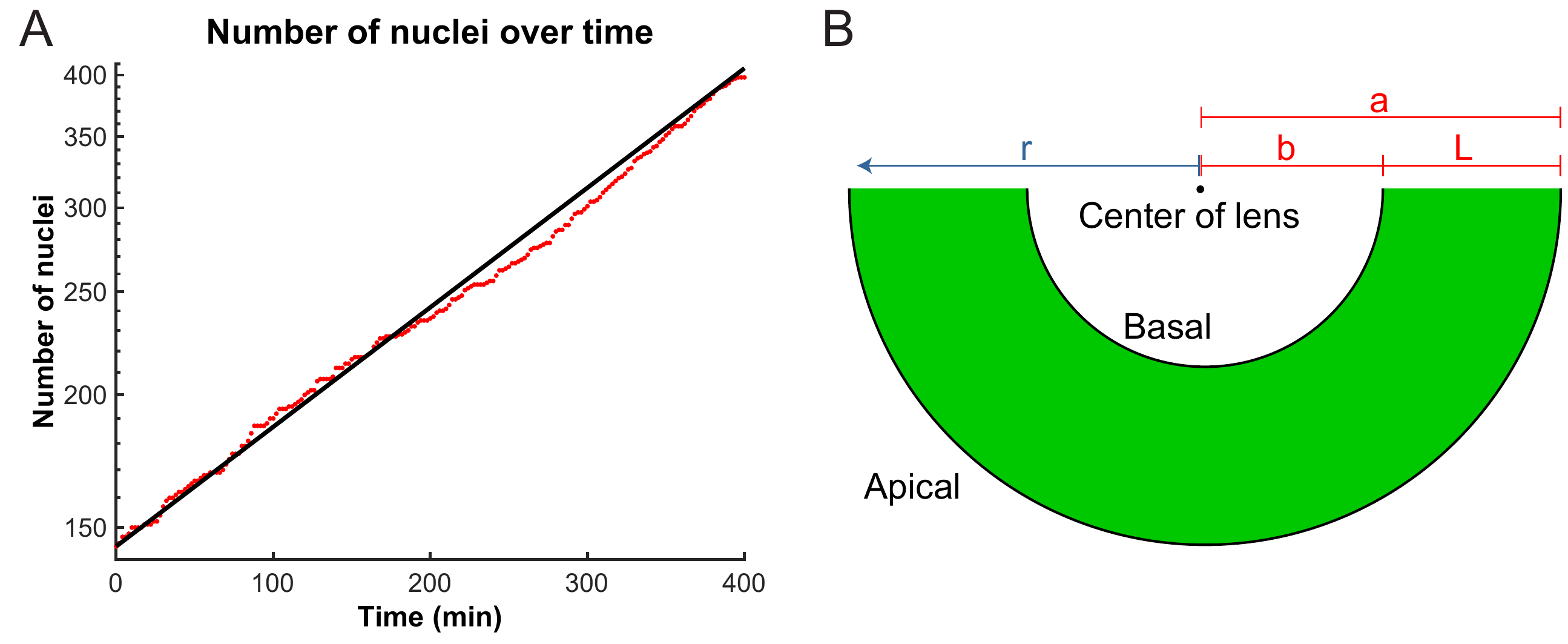}
\caption{\textbf{(A)} Number of nuclei grows exponentially during the proliferative stage of 
the retinal development. A line can be fit to the log-lin graph of nuclear numbers as a function 
of time to extract the doubling time (cell cycle length) in this period. \textbf{(B)} A schematic 
of the retina indicating the variables used in the diffusion model of IKNM. a: distance from 
center of lens to apical surface; b: distance from center of lens to basal surface; L: thickness 
of the retina; r: distance from center of lens for each particle.}
\label{fig:5}
\end{figure}

To investigate whether a diffusion model would also provide a useful quantitative description 
for IKNM, we formalized the process of IKNM in mathematical terms. This formalization again 
focuses on the crowding of nuclei at the apical side of the tissue. Crowding can be thought of, 
in mathematical terms, as creating a gradient in nuclear concentration $c$ along the apicobasal 
direction of the retina. In contrast, we assumed no dependence of the nuclear concentration 
on the lateral position within the tissue. Thus we employed the diffusion equation for the 
nuclear concentration $c(r,t)$ as a function only 
of the apicobasal distance 
$r$ and time $t$. The retina can be approximated as one half of a spherical shell around 
the lens, and thus we use spherical polar coordinates
with the origin of the coordinate system at the center of the lens, the basal surface at 
$r=b$ and the apical surface at $r=a$ (Figure~\ref{fig:5}B). We first study the simplest 
diffusion equation for this system,
in which there is a diffusion constant $D$ independent of position, time, and $c$ itself, namely
\begin{equation}
\frac{\partial c(r,t)}{\partial t}=D\frac{1}{r^2}\frac{\partial}{\partial r}
\left(r^2\frac{\partial c(r,t)}{\partial r}\right). \label{eqn_diffeqn}
\end{equation}
By analyzing the experimental data we seek to determine $D$. This equation provided the basis for our mathematical description of IKNM in terms of a diffusion process.

In addition to Equation~\ref{eqn_diffeqn}, we also needed to specify the boundary conditions 
adequate to describing IKNM. As mentioned above, we focused our description of IKNM on the 
apical crowding of nuclei. Since nuclei only divide close to the apical surface of the tissue, 
we treat mitosis as creating an effective influx of nuclei through the apical boundary. To quantify 
this influx, we extracted the number of cells $N(t)$ as a function of time. As during the stages 
of development examined here cells are neither dying nor exiting the cell cycle
\citep{Biehlmaier_Onset_2001}, we assumed that the number of cell divisions is always 
proportional to the number of currently existing cells. This assumption predicts an 
exponential increase in the number of cells or nuclei, over time, also recently found 
by \citet{Mateji_A_2018}:
\begin{equation} 
N(t) = N_0 e^{\nicefrac{t}{\tau}}, \label{eqn_growthofN}
\end{equation}
where $N_{0}$ is the initial number of nuclei and $\tau = T_P/\ln2$, with 
$T_{P}$ the average cell cycle length. Figure~\ref{fig:5}A shows the agreement between the 
theoretically predicted curve $N(t)$ with the experimentally obtained numbers of nuclei over 
time. Having obtained $N_{0}$ and $T_{P}$ from our experimental data, the predicted curve 
does not have any remaining free parameters and thus no fitting is necessary. Thus, the 
obtained description for the number of nuclei over time, Equation~\ref{eqn_growthofN}, was 
used to formulate the influx boundary condition for our mathematical model 
\begin{equation}
D\frac{\partial c}{\partial r}\Bigg\vert_{r=a} = \frac{1}{S} \frac{\partial N(t)}{\partial t} 
= \frac{N_0}{S\tau } e^{\nicefrac{t}{\tau}}, \label{eqn_apicalboundary}
\end{equation}
with $S$ the apical surface area of our domain of interest. 
In contrast to the apical side of the tissue, there is no creation (or depletion) of nuclei 
at the basal side \citep{Mateji_A_2018}, and hence a no-flux boundary condition,
\begin{equation}
\frac{\partial c}{\partial r}\Bigg\vert_{r=b}=0. \label{eqn_basalboundary}
\end{equation} 
Equations~\ref{eqn_diffeqn}, \ref{eqn_apicalboundary} and \ref{eqn_basalboundary} fully specify this simplest mathematical model of IKNM. 

From these equations we can derive an expression for the concentration of nuclei $c(r,t)$ in 
the retinal tissue. To this end, we introduced dimensionless variables for space and time,
\begin{equation}
\xi=\frac{r}{a}, \hspace{1cm} s=\frac{Dt}{a^2} \label{eqn_nondim}
\end{equation}
and further define $\rho=b/a<1$. The exact solution for the nuclear concentration, whose
detailed derivation is given in the Appendix, is
\begin{equation}
c(\xi,s) = \sum_{i=1}^{\infty} \left( h_i e^{- \lambda_i^2 s} 
+ \frac{\alpha_i f_0}{\sigma + \lambda_i^2} e^{\sigma s} \right) H_i(\xi) + \frac{1}{1-\rho} 
\left( \frac{1}{2} \xi^2 - \rho \xi + g_0\right) f_0 e^{\sigma s} \label{eqn_fullsolution}.
\end{equation}
The first terms within parentheses describes the decay over time of the initial 
condition $c_{\text{exp}}(\xi,s=0)$. Here, $\lambda_i$ are the eigenvalues and $H_i(\xi)$ 
the eigenfunctions of the radial diffusion problem, and the coefficients $h_i$ are 
determined from the experimental initial conditions (see Methods). The second terms 
within the sum and the final term on the right hand side of Equation \ref{eqn_fullsolution} 
are constructed such that the solution fulfills the boundary conditions \ref{eqn_apicalboundary} 
and \ref{eqn_basalboundary}. In the last term, the constant $g_0$ was obtained using the constraint 
that the volume integral of the initial concentration yields the initial number of nuclei $N_0$. 
$f_0$, $\sigma$ and $\alpha_i$ emerge within the calculation of the solution and are specified 
in the Appendix. Thus, the effective diffusion constant $D$ in Equations~\ref{eqn_diffeqn} 
and \ref{eqn_fullsolution} is the only unknown in the model.

\subsection{The linear model is accurate at early times}

\begin{figure}
\includegraphics[width=\linewidth]{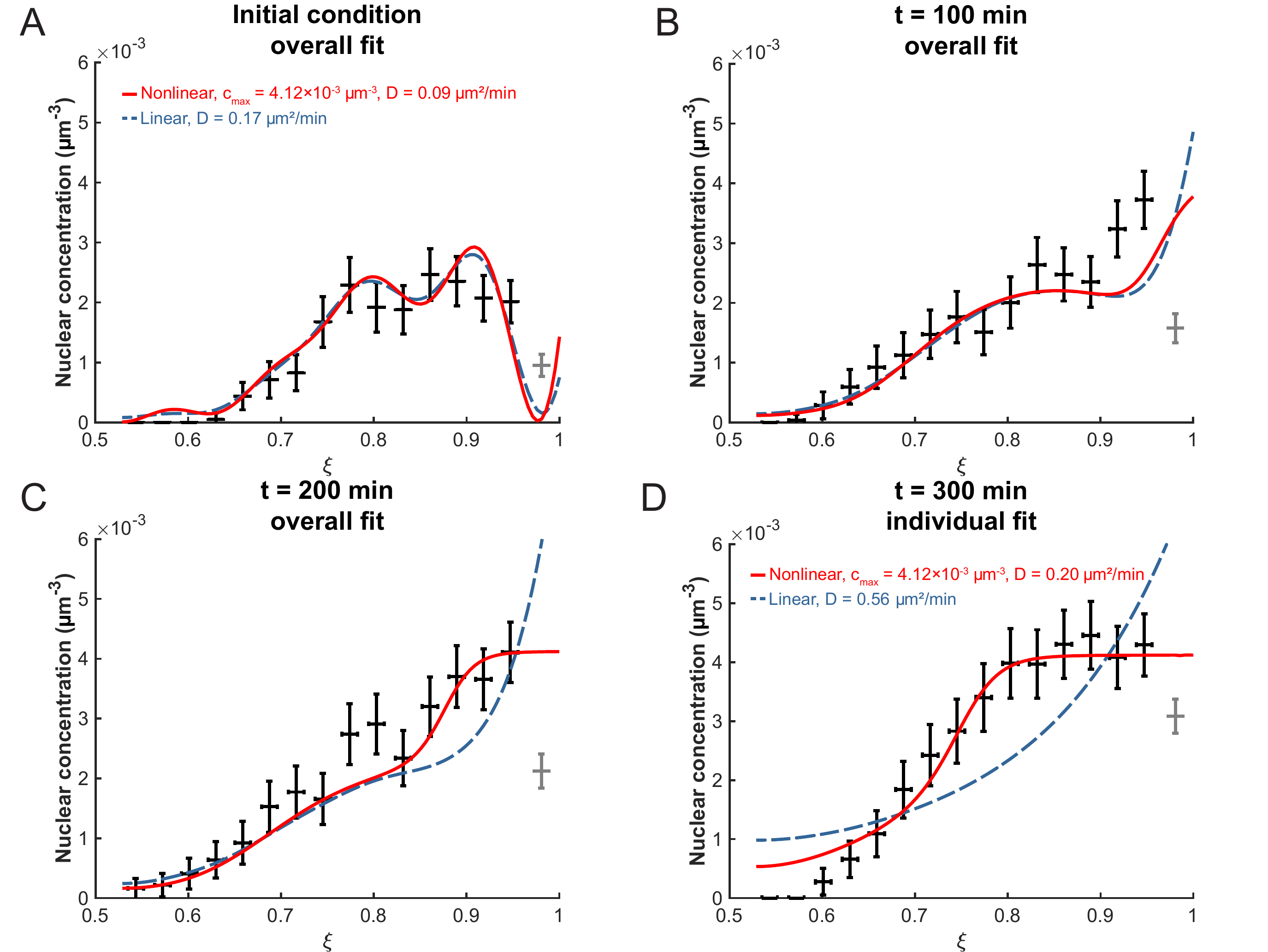}
\caption{\textbf{(A)} The initial experimental concentration profile of nuclei at $t = 0$ min 
as well as the calculated initial condition curves (see Methods Equation \ref{eqn_prefactors}) 
for the linear (red solid line) and nonlinear (blue dashed line) models. B,C,D) The fit of the 
models to experimental distribution of nuclei after 100 min \textbf{(B)}, 200 min \textbf{(C)}, and 
300 min \textbf{(B)} are shown. For the first three graphs, the best fit over all 100 intervening 
time points were used with the corresponding diffusion constants shown in \textbf{(A)}. For 
t = 300 min, the best fit at that time point only was used with the corresponding diffusion 
constants indicated.}
\label{fig:6}
\end{figure}

As mentioned before, the only parameter in the solution \ref{eqn_fullsolution} is the effective 
diffusion constant $D$. To determine this from the data, the experimentally obtained distribution 
of nuclei in the retinal tissue was first converted into a concentration profile. Then, the 
optimal $D$-value, henceforth termed $D^*$, was obtained using a minimal-$\chi^2$ approach. 
The value obtained within the linear model for a binning width of 3 \textmu m and an apical 
exclusion width of 4 \textmu m is $D^*_{\text{lin}} = 0.17 \pm 0.07$ \textmu m$^2$/min. 
Using this, we can examine the decay times of the different modes in the first term of 
Equation \ref{eqn_fullsolution}. The slowest decaying modes are the ones with the smallest eigenvalues 
$\lambda_i$ and we find that the longest three decay times are $\mathscr{T}_1 \approx 1325$ 
min, $\mathscr{T}_2 \approx 350$ min and $\mathscr{T}_3 \approx 158$ min. This shows that indeed 
all three terms of Equation \ref{eqn_fullsolution} are relevant on the timescale of our experiment 
and need to be taken into account when calculating the concentration profile. The corresponding 
plots of $c(\xi,s)$ are shown in Figure~\ref{fig:6}A-C. As can be seen from this figure, the 
diffusion model fits the data very well at early times, $t \leq 200$ min. However, for $t 
\geq 200$ min the model does not fit the data as well; the experimentally observed nuclear 
concentration levels off at a value between $4.00$ and $4.50 \times 10^{-3}$ \textmu m$^{-3}$
(Figure~\ref{fig:6}D), an aspect that is not captured by the model of linear diffusion. 

One particular aspect of the biology that the linear model neglects is the spatial extent 
of the nuclei. In a linear diffusion model, particles are treated as point-like and 
non-interacting. However, our microscopy images (see Figure \ref{fig:1}A) clearly indicate 
that the nuclei have finite incompressible volumes, so that their dense arrangement within 
the retinal tissue would lead to steric interactions once the nuclear concentration is 
sufficiently high, and moreover that the packing density of nuclei can not exceed a maximum 
value dictated by their geometry. Next, we examine whether accounting for these effects leads 
to a more accurate theory.

\subsection{Nonlinear extension to the model}

If we write the diffusion equation \ref{eqn_diffeqn} in the form
\begin{equation}
\frac{\partial c}{\partial t}=D\frac{1}{r^2}\frac{\partial}{\partial r}
\left\{r^2c\frac{\partial}{\partial r} \left[ \frac{\partial}{\partial c} 
\left( c \ln c \right) \right] \right\}, \label{eqn_diffeqn_rewritten}
\end{equation}
we can identify the term $c \ln c$ as proportional to the entropy ${\mathscr S}$ of an ideal gas, 
and its derivative with respect to $c$ as a chemical potential. In an ideal gas, all particles are 
treated as point-like and without mutual interactions. In order to include the spatial extent of
particles, we estimate the entropy using the model of a lattice gas, a system in which space is
divided into discrete sites which can either by empty or occupied by a single gas particle. Due 
to the discrete lattice, particles cannot get closer than the lattice spacing from each other, 
and there is a maximum possible concentration $c_{\text{max}}$ \citep{Huang_Statistical_1987}. 
In this system the entropy takes the form
\begin{equation}
{\mathscr S}_{\text{lattice gas}} \propto  c \ln c + \left( c_{\text{max}} - c \right) 
\ln \left( c_{\text{max}} - c \right).
\end{equation}
Substituting this expression for the term $c \ln c$ in \ref{eqn_diffeqn_rewritten}, we obtain
the nonlinear diffusion equation
\begin{equation}
\frac{\partial c}{\partial t} = D\frac{1}{r^2}\frac{\partial}{\partial r} 
\left( r^2 \frac{c_{\text{max}}}{c_{\text{max}} -c} \frac{\partial c}{\partial r} \right).
\label{eqn_diffeqn_nonlinear}
\end{equation}
Adjusting the boundary conditions at the apical side accordingly leads to
\begin{equation}
D \frac{c_{\text{max}}}{c_{\text{max}}-c} \frac{\partial c}{\partial r}\Bigg\vert_{r=a} = 
\frac{N_0}{S\tau} e^{\nicefrac{t}{\tau}}, \label{eqn_apicalboundary_nonlinear}   
\end{equation}
while the basal boundary condition remains the same as Equation~\ref{eqn_basalboundary}.
Together, Equation~\ref{eqn_diffeqn_nonlinear} and the boundary conditions in
Equations~\ref{eqn_apicalboundary_nonlinear} and \ref{eqn_basalboundary} represent an 
extension to the diffusion model for IKNM, which now accounts for steric interactions 
between the nuclei. The maximum concentration $c_{\text{max}}$ incorporated in this model 
was obtained, as described in the Methods, by considering a range of nuclear radii and the 
maximum possible packing density for aligned ellipsoids \citep{Donev_Unusually_2004}.

Similar to fitting the linear model, we also need to establish a description of the initial 
condition. To make both models consistent with each other, we employ the linear model's 
initial condition, Equation~\ref{eqn_fullsolution} at $s = 0$ with $h_i$ as obtained 
from Equation~\ref{eqn_prefactors}, as an initial condition for this nonlinear model as 
well (Figure \ref{fig:6}A). The concentration profile in the nonlinear model and its 
derivative were obtained numerically using the MATLAB pdepe solver. Fitting this 
concentration profile to the data was again by means of a minimal-$\chi^2$ approach. 
When the optimization took data points up to $t = 200$ min into account, we find 
$D^*_{\text{nonlin}} = 0.09 \pm 0.05$ \textmu m$^2$/min (Figure~\ref{fig:6}, Table \ref{tab:1}). 
As can be seen, by choosing $c_{\text{max}}$ correctly, an excellent fit to the data can be 
obtained. These results show that a lattice-gas based diffusion model is indeed suitable to 
describe time evolution of the nuclear concentration profile in zebrafish retina tissue 
during IKNM over several hours of development. 

\subsection{Incubation temperature has direct effects on IKNM}

The diffusion model may also address mechanistic questions about IKNM in retinas growing 
under varying experimental conditions. Zebrafish embryos are often grown at different 
temperatures to manipulate their growth rate \citep{Kimmel_Stages_1995,Reider_Effects_2014}, 
but it has been unclear how the nuclei in the retina behave at these different temperatures.  
To examine this issue, we grew the embryos at the normal temperature of 28.5~\si{\celsius} 
overnight and then incubated them at lower temperature (LT) of 25~\si{\celsius} or higher 
temperature (HT) of 32~\si{\celsius} during imaging. We could directly measure the change 
in average cell cycle length from experimental data and found that in HT, it is $205.5$ min, 
while in LT, it is a much longer $532.78$ min. We were then able to use these values in the 
model to investigate whether the change in temperature influences the processes that determine 
the effective diffusion constant of the nuclei. The resulting values for $D^*_{\text{nonlin}}$ are 
summarised in Table~\ref{tab:1}. Based on these values, two-sided \textit{t}-tests (see Methods) 
confirmed that there is no significant difference between the $D$-values obtained from the two 
normal condition data sets. In contrast, $D$-values for the LT and HT data sets were 
significantly different from the normal 
ones, with $p \leq 0.01$. These results indicate, that aside from its effect on cell 
cycle length, incubation temperature is likely to influence IKNM directly by altering 
the mobility of nuclei, here represented by the effective diffusion constant $D$. 

\begin{table}
\caption{List of best-fit diffusion constants $D^*$, their standard deviations and 
probabilities for the studied conditions.}
    \centering
    \begin{tabular}{l c c c}
    \toprule
        & $D^*_{\text{nonlin}}$ (\textmu m$^2$/min) & $\sigma_{D}$ (\textmu m$^2$/min) &
        $P_{\chi}(\chi^2;\nu)$ \\
    \midrule
       Normal  &  0.09 & 0.05 & 0.49 - 0.51 \\
       Normal (repeat sample) & 0.10 & 0.06 & 0.47 - 0.48 \\
       High T & 0.13 & 0.08 & 0.42 \\
       Low T & 0.06 & 0.05 & 0.69 - 0.7 \\
    \bottomrule
    \end{tabular}
    \label{tab:1}
\end{table}

\section{Discussion}

In this work, we have shown that high density nuclear trajectories can be used to tease apart 
the possible physical processes behind the apparently stochastic movement of nuclei during 
interkinetic nuclear migration. Firstly, we generated these trajectories using long-term 
imaging and tracking of nuclei with high spatial and temporal resolution within a 
3-dimensional segment of the zebrafish retina. Analysis of speed and positional distributions 
of more than a hundred nuclei revealed a large degree of variability in their movements during 
G1 and S phases. Although this variability had been observed before, previous experiments had 
only considered sparsely labeled nuclei within an otherwise unlabeled environment
\citep{Baye_Interkinetic_2007,Norden_Actomyosin_2009,Leung_Apical_2011}. Thus, our results 
provide an important account of the variability of IKNM on a whole tissue level. In effect, 
the variability of IKNM means that nuclear trajectories appear stochastic during the majority 
of the cell cycle. Previously, it had been suggested that  the origins of this apparent 
stochasticity lay in the transfer of kinetic energy between nuclei in G2 exhibiting rapid 
apical migration to nuclei in G1 and S phases of the cell cycle, much as a person with an 
empty beer glass may nudge away other customers to get to the bar \citep{Norden_Actomyosin_2009}. 
However, we found no evidence for direct transfer of kinetic energy between nuclei and their 
immediate neighbors. Recently \citet{Shinoda_Elasticity_2018} have also provided evidence that 
suggests direct collisions do not contribute to basal IKNM. 

Another possibility is that the stochastic trajectories of G1 and S nuclei could be a result 
of passive displacements, arising from a diffusive process depending on a nuclear concentration 
gradient between the apical and basal sides of the tissue \citep{Miyata_Interkinetic_2015}. This 
gradient could be formed by nuclear divisions taking place exclusively at the apical surface. We 
confirmed the presence of such a gradient by calculating the nuclear concentration along the 
apicobasal dimension within the retinal tissue at various time points. Further, to probe the 
source of the gradient, we treated the zebrafish retina with HU-AC to stop the cell cycle in 
S phase. While we observed the build-up of the nuclear concentration gradient over time in 
the control retina, the nuclear distribution flattened when cell division was inhibited with 
HU-AC treatment. These phenomenological similarities between IKNM and diffusion suggested the 
diffusive model. This model includes two key features: most importantly, it focuses on the 
crowding of nuclei at the apical surface of the tissue, here included as the apical boundary 
condition. Additionally, in the nonlinear extension of the model, it incorporates a maximum 
possible nuclear concentration. This addition provided a striking overall improvement to the 
fits to experimental data over periods of many hours. The resulting difference in the obtained 
$D$-values between the linear and nonlinear versions of our model can be understood heuristically 
when closely examining the difference between Eqs. \ref{eqn_diffeqn} and \ref{eqn_diffeqn_nonlinear}. 
The latter introduces the new term $c_{\text{max}}/(c_{\text{max}} -c)$ which one could think 
of loosely as corresponding to an effective, concentration dependent diffusion constant 
$\tilde D = Dc_{\text{max}}/(c_{\text{max}} -c)$. In general $\tilde D$ will vary across 
the tissue thickness and, since $c > 0$ for most of the retinal tissue, $\tilde D > D$. 
Therefore, averaging across the retina tissue, $\tilde D$ might actually be in very good 
agreement with the $D$-value found in the linear model. However, the fact the linear model 
fails to describe, and which leads to a better representation of the data using the nonlinear 
model, is that the mobility of the nuclei is likely to be concentration dependent. 

The underlying processes causing IKNM during the G1 and S phases of the cell cycle in 
pseudostratified epithelia have been largely elusive. Several partially competing ideas 
have been put forward, ranging from the active involvement of cytoskeletal transport 
processes to passive mechanisms of direct energy transfer or movements driven by apical 
nuclear crowding \citep{Schenk_Myosin_2009,Tsai_Kinesin_2010,Norden_Actomyosin_2009,Kosodo_Regulation_2011}. 
The fact that inanimate microbeads migrate much like nuclei during IKNM in the mouse cerebral 
cortex \citep{Kosodo_Regulation_2011} suggests that active, unidirectional intracellular 
transport mechanisms are not directly responsible for these stochastic movements. Instead, 
we showed that a passive diffusive process which takes steric interactions between nuclei 
into account produces an excellent representation of the time evolution of the actual 
nuclear distribution within the retinal tissue during early development. Consequently, 
our work builds on earlier models of apical crowding based on \textit{in silico} simulations 
of IKNM \citep{Kosodo_Regulation_2011}. Having said this, it remains 
to understand the general scale of the diffusion constant ($D\sim$0.1
$\mu^2$/min) from microscopic considerations, perhaps analogous to
those used to relate random walks to diffusion \citep{Goldstein2018}.
In addition, our work revealed the remarkable importance of simple physical constraints 
imposed by the overall tissue architecture, which could not be  explored in previous studies 
which tracked sparse nuclei, and thus lacked the means to explore the effect of such 
3-dimensional arrangements. Hence, we paid special attention to the spherical shape of 
the retina and the concentration of nuclei in that space. Examining the evolution in 
distribution of nuclei over time unveils the importance of spatial restriction due to 
the curvature of the tissue. Additionally, the size of the nuclei in comparison to the 
neuroprogenitor cells leads to the emergence of a maximum nuclear concentration which 
must be taken into account to accurately model IKNM.

By inhibiting cell cycle progression or changing temperature, we used  our model to shed some 
light on some of the properties of and mechanisms of the stochastic movements of nuclei during 
IKNM. From our results and previous studies, we knew that cell cycle length is affected by 
change in incubation temperature \citep{Kimmel_Stages_1995,Reider_Effects_2014}. However, our 
results also indicate a significant influence of temperature on the mobility of nuclei and 
thus the underlying processes controlling their movement. For example, the speed and dynamic 
properties of both the microtubule and actomyosin systems are dependent on temperature and 
could in part explain the changes in the diffusion constant that we see as a function of 
temperature \citep{Hartshorne_the_1972,hong_the_2016} as the diffusion constant may be 
influenced by stochastic associations with motor proteins or the physical properties of the 
epithelium.  However, a much closer examination of molecular mechanisms driving stochastic 
nuclear movements is required to better understand the connections between these phenomena, 
as we are far from understanding the nature of forces involved in this process.  Furthermore, 
the diffusion constant reported here contains all types of nuclear movement during IKNM as it 
is derived from the changing nuclear concentration profile over time. However, it is not 
immediately clear what the contribution the rapid apical migration to this overall diffusion 
constant may be. Nonetheless, despite the large displacement during rapid apical migration at 
G2, this phase only accounts for about 8\% of the cell cycle \citep{Leung_Apical_2011}. 
Therefore, given this small portion of the cell cycle when rapid migration can happen and 
the good agreement of our calculated diffusion constant with those previously reported in 
the literature for individual nuclei \citep{Leung_Apical_2011}, the proposed model appears 
to describe tissue-wide IKNM quite well.

The physiological consequences of nuclear arrangements and the IKNM movements associated 
with all pseudostratified epithelia are not well understood. Our results provide a quantitative
description of the stochastic distribution of the nuclei across the retina. This distribution 
has been implicated in stochastic cell fate decision making of progenitor cells during 
differentiation \citep{Clark_Loss_2012,Baye_Interkinetic_2007,Hiscock_Feedback_2018}. Our 
observations would fit with previous suggestions that a signalling gradient, such as a Notch 
gradient, exists across the retina and location-dependent exposure to it is important for 
downstream decision-making \citep{Murciano_Interkinetic_2002,Del_Bene_Regulation_2008,Hiscock_Feedback_2018,Aggarwal_Concentration_2016}. Thus, our results not only have important implications for 
understanding the organisation of developing vertebrate tissues, but may also provide a starting 
point for further exploration of the connection between variability in nuclear positions and 
cell fate decision making in neuroepithelia.

\section{Methods and Materials}

\subsection{Animals and Transgenic Lines}
All animal work was approved by Local Ethical Review Committee of the University of Cambridge 
and performed in accordance with a Home Office project license PL80/2198.
All zebrafish were maintained and bred at 26.5~\si{\celsius}. All embryos were incubated 
at 28.5~\si{\celsius} before imaging sessions. At 10 hours post fertilization (hpf), 0.003\%
phenylthiourea (PTU) (sigma) was added to the medium to stop pigmentation in the eye. 

\subsection{Lightsheet microscopy}
Images of retinal development for the main dataset were obtained using lightsheet microscopy. 
Double transgenic embryos, Tg(bactin2:H2B-GFP::ptf1a:DsRed) were dechorionated at 24 hpf and 
screened positive for the fluorescent transgenic markers prior to the imaging experiment. The 
embryo selected for imaging was then embedded in 0.4\% low gelling temperature agarose 
(Type VII, Sigma-Aldrich) prepared in the imaging buffer (0.3x Daniau’s solution with 0.2\% 
tricaine and 0.003\% PTU \citep{Godinho_imaging_2011}) within an FEP tube with 25 \textmu m 
thick walls (Zeus), with an eye facing the camera and the illumination light shedding from 
the ventral side. The tube was held in place by a custom-designed glass capillary (3 mm outer 
diameter, 20 mm length; Hilgenberg GmbH). The capillary itself was mounted vertically in the 
imaging specimen chamber filled with the imaging buffer. To ensure normal development, a 
perfusion system was used to pump warm water into the specimen chamber, maintaining a 
constant temperature of 28.5~\si{\celsius} at the location of the specimen. 

Time-lapse recording of retinal development was performed using a SiMView light-sheet 
microscope \citep{Tomer_Quantitative_2012} with one illumination and one detection arm. 
Lasers were focused by Nikon 10x/0.3 NA water immersion objectives. Images were acquired 
with Nikon 40x/0.8 NA water immersion objective and Hamamatsu Ocra Flash 4.0 sCMOS camera. 
GFP was excited with scanned light sheets using a 488 nm laser, and detected through a 525/50 
nm band pass detection filter (Semrock). Image stacks were acquired with confocal slit 
detection \citep{Baumgart_Scanned_2012} with exposure time of 10 ms per frame, and the sample 
was moved in 0.812 \textmu m steps along the axial direction. For each time point, 
two 330 x 330 x 250 \textmu m$^3$ image stacks with a 40 \textmu m horizontal offset 
were acquired to ensure the coverage of the entire retina. The images were acquired every 
2 min from 30 hpf to 72 hpf. The position of the sample was manually adjusted during imaging 
to compensate for drift. The two image stacks in the same time point were fused together to 
keep the combined image with the best resolution. An algorithm based on phase correlation 
was subsequently used to estimate and correct for the sample drift over time. The processing 
pipeline was implemented with MATLAB (MathWorks).

\subsection{Two photon microscopy}
Images for the repetition dataset and all other conditions were obtained using a TriM Scope 
II 2-photon microscope (LaVision BioTec). A previously established Tg(H2B-GFP) line, generated 
by injecting a DNA construct of H2B-GFP driven from the actin promoter \citep{He_How_2012}, 
was used for all these experiments. Embryos were dechorionated and screened for expression of 
GFP at 24 hpf. An embryo was then embedded in 0.9\% UltraPure low melting point agarose 
(Invitrogen) prepared in E3 medium containing 0.003\% PTU and 0.2\% tricaine. The agarose 
and embryo were placed laterally within a 3D printed half cylinder of transparent ABS plastic, 
0.8 mm in diameter, attached to the bottom of a petri dish, such that one eye faced the 
detection lens of the microscope. The petri dish was then filled with an incubation solution 
of E3 medium, PTU, and tricaine in the same concentrations as above. For the experiment 
involving cell cycle arrest, hydroxyurea and aphidicolin (Abcam) were added to the incubation 
solution right before imaging, to a final concentration of 20 mM and 150 \textmu M, respectively. 
The imaging chamber was maintained at a temperature of 25~\si{\celsius}, 28.5~\si{\celsius}, 
or 32~\si{\celsius}, as required, using a precision air heater (The Cube, Life Imaging Services).

Green fluorescence was excited using an Insight DeepSee laser (Spectra-Physics) at 927 nm. 
The emission of the fluorophore was detected through an Olympus 25x/1.05 NA water immersion 
objective, and all the signal within the visible spectrum was recorded by a sensitive GaAsP 
detector. Image stacks with step size of 1 \textmu m were acquired with exposure time of 
1.35 ms per line averaged over two scans. The images were recorded every 2 min for 10-15 hours 
starting at 26-28 hpf. The same post processing procedure for data compression and drift correction 
was used on these raw images as on those from lightsheet imaging.

\subsection{Obtaining experimental input values for the model}

The radial coordinates $r_n$ of nuclei were calculated by subtracting $l_n$ from $a$, wherein 
$l_n$ is the distance from the center of a nucleus $n$ to the apical surface and $a$ is the 
distance from the center of the lens to the apical surface. We estimated a total uncertainty of 
$\Delta r = \pm 3$ \textmu m for each single distance measurement of $r_n$. This value is a 
result of uncertainty in detecting the center of the nucleus and in establishing the position 
of the apical surface.

Because each nuclear position has an error bar $\Delta r$, binning the data leads to an 
uncertainty in the bin count. In order to calculate this uncertainty, we considered the 
probability distribution of a nucleus' position. In the simplest case, this probability 
is uniform within the width of the positional error bar and zero elsewhere. The probability,
$p_{n,\text{bin}}$, of finding a given nucleus $n$ within a given bin, is proportional to the 
size of the overlap of probability distribution and bin. It follows that the expectation value 
for the number of nuclei within a bin is given as $\mathbb{E}(N_{\text{bin}}) = 
\sum_n p_{n,\text{bin}}$. Correspondingly, $\text{Var}(N_{\text{bin}}) = \sum_n p_{n,\text{bin}}
(1-p_{n,\text{bin}})$ is the variance of the number of nuclei within this bin. Thus, the 
error bar of the bin count is $\sigma_{y,\text{bin}} = \sqrt{\text{Var}(N_{\text{bin}})}$. 
The nuclear distribution profile $N(r,t)$ is not expected to be uniform or linear, therefore 
the expectation value $\mathbb{E}(N_{\text{bin}})$ does not correspond to the number of nuclei 
at the center of the bin. Since the position of the expectation value is unknown \textit{a priori}, 
it is still plotted at the center of the bin with an error bar denoting its positional uncertainty. 
Here we assume this error bar to be the square-root of the bin size $\Delta r_{\text{bin}}$, 
i.e. $\sigma_{x,\text{bin}} = \sqrt{\Delta r_{\text{bin}}}$.

In order to obtain the experimental nuclear concentration profile $c(r,t)$, and its error bars, 
from the distribution of nuclei $N(r,t)$, the volume of the retina also has to be taken into 
account, since $c=N/V$. The total retinal volume within which nuclei tracking took place was 
estimated directly from the microscopy images. To this end, we outlined the area of observation 
in each image slice using the Fiji software and multiplied this area with the distance between 
successive images. Given the total volume, $V_{\text{total}}$, we proceeded to calculate the 
volume per bin, which depends on the radii at the inner and outer bin surfaces. In general, 
the volume of part of a sphere, e.g. a spherical sector, is given as $V_{\text{sector}} =
\frac{1}{3}\Omega r_{\text{sector}}^3$, where $\Omega$ denotes the solid angle. Knowing the 
apical and basal tissue radii, $r=a$ and $r=b$, one can thus calculate $\Omega$ as 
$\Omega =3V_{\text{total}}/(a^3-b^3)$. This gives the volume of each bin as $V_{\text{bin}} =
\frac{1}{3}\Omega \left(r_{\text{bin,outer}}^3 - r_{\text{bin,inner}}^3 \right)$, where
$r_{\text{bin,outer}}$ and $r_{\text{bin,inner}}$ denote the outer and inner radii of a bin,
respectively. Similarly, we calculated the effective surface area $S$ through which the influx 
of nuclei occurs (see Equation~\ref{eqn_apicalboundary}) from the solid angle $\Omega$. This 
surface area is simply given as $S = \Omega a^2$. 

To retrieve the average cell cycle time $T_P$ for each of the data sets, we used two different
approaches. In the case of the main data set, sufficient number of nuclear tracks consisting 
of a whole cell cycle were present. Thus we directly calculated the average cell cycle duration 
from these tracks. For the other datasets, we make use of the fact that the number of nuclei 
follows an exponential growth law depending on $T_P$ (see Equation~\ref{eqn_growthofN}). 
Knowing the initial number of tracked nuclei $N_0$ for each data set, we obtained $T_P$ from 
fitting the following equation to the number of nuclei as a function of time in a log-lin plot: 
$\ln N(t) = \ln N_0 + t/\tau = \ln N_0 + (\ln 2/T_P) t$. Then $T_P$ was 
deduced from the slope of this fit.

In order to determine the maximum nuclear concentration $c_{\text{max}}$ for the nonlinear 
model, we first randomly selected 100 nuclei from our dataset of tracked nuclei and measured 
the size of their longest diameter in both XY and YZ planes. From these measurements we 
established that the size of the principal semi-axis of each nucleus is likely to lie 
in the range of about 3 \textmu m to 5 \textmu m, where the nuclear shape is regarded to 
be ellipsoidal. This led to the range of possible maximum concentrations $c_{\text{max}}$, 
although we did not measure the precise nuclear volume. The lower limit for the nuclear 
volume is set by the volume of a sphere of radius 3 \textmu m, the upper limit by a sphere 
of radius 5 \textmu m. Taking into account the maximum possible packing density of nuclei, 
which for aligned ellipsoids is the same as that of spheres \citep{Donev_Unusually_2004}, 
$\frac{\pi}{3 \sqrt{2}} \approx 0.74$, we obtained a range of $1.41 
\times 10^{-3}$ \textmu m$^{-3}$ $\leq c_{\text{max}} \leq 6.55 \times 10^{-3}$ \textmu m$^{-3}$.  

\subsection{Obtaining the initial condition}

We determined the prefactors $h_i$ from the experimental nuclear distribution at the start of 
the experiment, $c_{\text{exp}}(\xi,0)$. For convenience, we chose to determine first 
$\widetilde{h_i} = h_i + \alpha_i f_0/(\sigma+\lambda_i^2)$ and then obtained $h_i$ by subtracting
$\alpha_i f_0/(\sigma+\lambda_i^2)$ from the results. The $\widetilde{h_i}$ can be calculated from 
the data, using Equation~\ref{eqn_fullsolution} for $s = 0$, as
\begin{equation}
    \widetilde{h_i} = \sum_m \xi_m^2 H_i(\xi_m) c_{\text{exp}}(\xi_m,0) \Delta\xi_{\text{m}} 
    - \frac{f_0}{1-\rho} \int_{\rho}^1 \xi^2 H_i(\xi) \left( \frac{1}{2}\xi^2 -\rho \xi + 
    g_0 \right) d\xi, \label{eqn_prefactors}
\end{equation}
where $m$ denotes the $m$-th binned data point, $\xi_m$ its position and $\Delta \xi_{\text{m}}$ 
the width of bin $m$. As in Equation~\ref{eqn_fullsolution}, the index $i$ denotes the $i$-th
eigenfunction or -mode.

\subsection{The concentration profile in the nonlinear model}

The non-linear concentration profile was determined numerically from the same initial 
condition as used for the linear model, Equation \ref{eqn_fullsolution}, at $s=0$ 
with $\widetilde{h_i}$ as in Equation \ref{eqn_prefactors}. Time evolution of the initial 
condition, according to Equation \ref{eqn_diffeqn_nonlinear}, was performed using the pdepe 
solver in MATLAB. 

\subsection{Fitting the model}

The range of sizes of the nuclear principal semi-axes was used to determine the range of 
data to be included in our fits. Any data closer than 3 \textmu m to 5 \textmu m from the 
apical or basal tissue surfaces was not taken into account for fitting because the center 
of a nucleus cannot be any closer to a surface than the nuclear radius. Thus, all data 
collection very close to the apical or basal tissue surfaces must have been due to the 
above mentioned measurement uncertainties $\Delta r$. 

In principle, the full solution for $c(\xi,s)$ is composed of infinitely many modes. 
However, in practice, we truncated this series and only included the first 8 modes in our fits. 
This is due to the fact that we have a finite set of data points, so adding too many modes could 
lead to over-fitting. Fits with a wide range of numbers of modes were found to result in the 
same optimal $D$-values.  

For fitting, we first rescaled the data in accordance with the non-dimensionalisation of the 
theoretical variables $r$ and $t$ (see Equation~\ref{eqn_nondim}). Thus we obtain 
$c_{\text{exp}}(\xi,s)$ from $c_{\text{exp}}(r,t)$. Then both models were fitted to the 
experimental data using a minimal-$\chi^2$ approach. The goodness of fit parameter $\chi^2 = 
\sum_m \left( c_{\text{exp}}(\xi,s) - c(\xi,s) \right)^2/\sigma_m^2$, where 
$\sum_m$ denotes the summation over all bins $m$.  
Since binning resulted in uncertainties $\sigma_{y,\text{bin}}$ and $\sigma_{x,\text{bin}}$ in the 
$y$- and $x$-directions, both had to be taken into account when calculating $\sigma_m$ and $\chi^2$. 
The combined contribution of $x$- and $y$- uncertainties is: $\sigma_{m}^2 = \sigma_{y,m}^2 +
\sigma_{y,\text{indirect},m}^2$ with  $\sigma_{y,\text{indirect},m} = \sigma_{x, m} 
\left(\mathrm d c(\xi,s)/\mathrm d \xi\right) \big\vert_{\xi=\xi_m}$ 
\citep{Bevington_Data_2003}. In our fits, the value $\chi^2$ was calculated for a large 
range of possible diffusion constants $D$, from $D=0.01$ \textmu m$^2$/min to $D=10$ 
\textmu m$^2$/min. By finding the value of $D$ for which $\chi^2$ became minimal for a 
given data set and time point, we established our optimal fit. 

The minimal-$\chi^2$ approach furthermore enabled us to determine the optimal binning 
width $\Delta r_{\text{bin}}$ or $\Delta \xi_{\text{bin}}$ and width of data exclusion 
for the fits. In order to do so, fits of the normal data set were performed for different 
data binning widths and exclusion sizes of 3 \textmu m to 5 \textmu m. For each of these 
fits the $\chi^2$-value and the number of degrees of freedom $\nu$, i.e.\ the number of data 
points minus the number of free fit parameters (here number of data points minus 1), were 
registered. From $\chi^2$ and $\nu$ we calculated the reduced $\chi^2$ value, $\chi_\nu^2 = 
\chi^2/\nu$ \citep{Bevington_Data_2003}. Using $\nu$ and $\chi_\nu^2$, the probability $P_\chi(\chi^2;\nu)$ of exceeding $\chi$ for a given fit can be estimated, which should be 
approximately 0.5 \citep{Bevington_Data_2003}. Therefore, we found our optimal data binning 
width of 3 \textmu m to 4 \textmu m as the width that resulted in a $P_\chi(\chi^2;\nu)$ as 
close to 0.5 as possible for all the different time points when fitting the nonlinear model. 
The exact choice of exclusion width was found not to influence the fitting result for the 
nonlinear model.

In addition to finding the optimal $D$-value for individual time points, we also modified the
minimal-$\chi^2$ routine to find the value of $D$ that fits a whole data set (i.e.\ all time 
points simultaneously) in the best possible way. In order to do so, we summed the $\chi^2$-values
obtained for each $D$ over all time points, in this way producing a $\sum_t\chi^2(D)$-curve. The 
minimum of this curve indicates $D^*$ for the whole time series. Furthermore, dividing 
$\sum_t\chi^2(D)$ by the number of time points included in the optimization yields an average 
$\chi^2$- and reduced $\chi^2$-value corresponding to this $D^*$. In addition, the width of this 
time averaged curve at $\chi^2 = \chi^2_{\text{min}} + 1$ indicates the standard deviation of the 
optimal $D$-value, $\sigma_D$. By approximating the minimum with a quadratic curve, we obtain an 
estimate for this standard deviation as $\sigma_D = \Delta_D \sqrt{2\left( \chi^2_{D^*-\Delta_D}
-2\chi^2_{D^*} + \chi^2_{D^*+\Delta_D} \right)}$ \citep{Bevington_Data_2003}  where $\Delta_D$ is 
the step size between individual fitted $D$-values, here $\Delta_D = 0.01$ \textmu m$^2$/min. 
Lastly, based on the average reduced $\chi^2$-values, we also compared several $c_{\text{max}}$-values
for each data set to find the fit with probability $P_\chi(\chi^2;\nu)$ the closest to 0.5 in each case. 

All fits were performed using custom MATLAB routines. 

\subsection{\textit{t}-tests}

To compare results between data sets, the values $D^*$ and corresponding $\sigma_D$ from the 
overall fits were considered. It should be noted that these values were not obtained by averaging 
several data sets of the same experimental condition but instead each value results from one data 
set only. However, the sample size for each data set was set to 100 because 100 time points were 
taken into account for each overall optimization. These time points might not be completely 
uncorrelated, limiting the predictive power of the \textit{t}-test. Two sided tests, specifically 
unequal variances \textit{t}-test, also known as Welch's \textit{t}-test, \citep{Precht_Bio_2015}, 
were performed in order to determine whether samples differ significantly from each other.

\section{Acknowledgments}

AH and REG would like to thank Oliver Y. Feng, Timothy J. Pedley, Michael E. Cates and Salvatore 
Torquato for helpful advice and input. This work was supported by the Cambridge Wellcome Trust PhD Programme in Developmental Biology, the Cambridge Commonwealth, European and International Trust, and Natural Sciences and Engineering Research Council of Canada (AA); the Engineering and Physical Sciences Research Council and a Helen Stone Scholarship at the University of Cambridge through the Cambridge Trust (AH); Established Career Fellowship EP/M017982/1 from the Engineering and Physical Sciences Research Council (REG); and Wellcome Trust Investigator Award (SIA 100329/Z/12/Z) (WAH).

\bibliography{IKNM}


\section{Appendix}

\subsection{Full solution of the linear diffusion equation}

After rescaling space and time as in Equation \ref{eqn_nondim} and introducing 
$\rho = a/b < 1$, Equation \ref{eqn_diffeqn} and the boundary conditions 
\ref{eqn_apicalboundary} and \ref{eqn_basalboundary} read
\begin{equation}
\begin{split}
    & \frac{\partial c(\xi,s)}{\partial s} = \frac{1}{\xi^2} \frac{\partial}{\partial \xi} 
    \left( \xi^2 \frac{\partial c(\xi,s)}{\partial \xi} \right), \\
    & \frac{\partial c(\xi,s)}{\partial \xi}\Bigg\vert_{\xi = 1} = f_{0} e^{\sigma s} = 
    f(s) \hspace{1cm} \text{and} \hspace{1cm} \frac{\partial c(\xi,s)}{\partial \xi} 
    \Bigg\vert_{\xi = \rho} = 0,
\end{split}
\end{equation}
where we have defined $f_{0} = aN_0/DS\tau$ and $\sigma = a^2/D\tau$.
We transform this homogeneous differential equation with inhomogeneous boundary conditions 
into the problem of solving an inhomogeneous differential equation with homogeneous boundary 
conditions by writing $c(\xi,s)$ as a sum of two contributions,
\begin{equation}
    c(\xi,s) = \phi(\xi,s) + \psi(\xi,s),
\end{equation}
where we require $\phi(\xi,s)$ to satisfy the inhomogeneous boundary conditions
\begin{equation}
     \frac{\partial \phi(\xi,s)}{\partial \xi}\Bigg\vert_{\xi = 1} = f_{0} e^{\sigma s}  
     \hspace{1cm} \text{and} \hspace{1cm} \frac{\partial \phi(\xi,s)}{\partial \xi} 
     \Bigg\vert_{\xi = \rho} = 0.
\end{equation}
These conditions are satisfied if $\phi(\xi,s)$ has the form
\begin{equation}
    \phi(\xi,s) = \frac{1}{1-\rho} \left( \frac{1}{2} \xi^2 - \rho \xi + g_{0} \right)f_{0} 
    e^{\sigma s}.
\end{equation}
where $g_{0}$ is a constant of integration to be determined later.
The remaining problem to solve for $\psi(\xi,s)$ is
\begin{equation}
    \frac{\partial \psi(\xi,s)}{\partial s} = \frac{1}{\xi^2} \frac{\partial}{\partial \xi} 
    \left( \xi^2 \frac{\partial \psi(\xi,s)}{\partial \xi} \right) + \frac{f_{0} e^{\sigma s}}{1-\rho}
    \left( 3 - \frac{2\rho}{\xi} - \sigma \left( \frac{1}{2} \xi^2 - \rho \xi + g_{0} \right) \right),
    \end{equation}
with homogeneous boundary conditions
\begin{equation}
     \frac{\partial \psi(\xi,s)}{\partial \xi}\Bigg\vert_{\xi = 1} = 0  \hspace{1cm} \text{and}
    \hspace{1cm} \frac{\partial \psi(\xi,s)}{\partial \xi}\Bigg\vert_{\xi = \rho} = 0. \label{eqn_forpsi}
\end{equation}
We can further write $\psi(\xi,s)$ as the sum of two contributions,
\begin{equation}
    \psi(\xi,s) = \psi_{h}(\xi,s) + \psi_{p}(\xi,s),
\end{equation}
where $\psi_{h}$ is the general solution of the homogeneous problem 
\begin{equation}
\begin{split}
    & \frac{\partial \psi_{h}(\xi,s)}{\partial s} = \frac{1}{\xi^2} \frac{\partial}{\partial \xi} 
    \left( \xi^2 \frac{\partial \psi_{h}(\xi,s)}{\partial \xi} \right), \\
    & \frac{\partial \psi_{h}(\xi,s)}{\partial \xi}\Bigg\vert_{\xi = 1} = 0  \hspace{1cm} 
    \text{and} \hspace{1cm} \frac{\partial \psi_{h}(\xi,s)}{\partial \xi} \Bigg\vert_{\xi = \rho} = 0,
    \label{eqn_forpsih}
    \end{split}
\end{equation}
and $\psi_{p}$ is a particular solution of the full inhomogeneous problem \ref{eqn_forpsi}.
The full solution of the homogeneous problem is given as a series of linearly independent 
eigenfunctions, each of the form
\begin{equation}
    e^{-\lambda^2 s}W(\xi) = e^{-\lambda^2 s} \left( A \frac{\sin{\lambda \xi}}{\xi} + 
    B \frac{\cos{\lambda \xi}}{\xi} \right), \label{eqn_eigfctnonnormal}
\end{equation}
where the eigenvalues $\lambda$ can be found from simultaneous solution of the boundary conditions,
\begin{equation}
    \begin{split}
        & A\left(\lambda \cos{\lambda}-\sin{\lambda}\right)- B \left(\lambda \sin{\lambda} + \cos{\lambda}\right) = 0 \\
        & A \left(\frac{\lambda \cos{\lambda \rho}}{\rho} - \frac{\sin{\lambda \rho}}{\rho^2}\right) - 
        B\left(\frac{\lambda \sin{\lambda \rho}}{\rho}+ \frac{\cos{\lambda \rho}}{\rho^2}\right) = 0,
        \label{eqn_boundariesfortranscendental}
    \end{split} 
\end{equation}
which yields the transcendental relation
\begin{equation}
    \tan{\lambda \left( 1-\rho \right)} = \frac{\lambda \left( 1-\rho \right)}{\lambda^2 \rho +1}, \label{eqn_transcendental}
\end{equation}
for which each eigenvalue $\lambda_{i}$ is a solution corresponding 
to one of the linearly independent eigenfunctions (only $\lambda_{i} > 0$ need to be 
taken into account).
We can further deduce from the Equation \ref{eqn_boundariesfortranscendental} that
$B_i=\beta_i A_i$, where
\begin{equation}
     \beta_{i} = \frac{\lambda_{i} \cos{\lambda_{i}} - \sin{\lambda_{i}}}{\lambda_{i} 
    \sin{\lambda_{i}} + \cos{\lambda_{i}}},
\end{equation}
and we normalize the obtained expression for $W_{i}(\xi)$ from Equation \ref{eqn_eigfctnonnormal}
\begin{equation}
    H_{i}(\xi) = \frac{1}{Y_{i}} \left( \frac{\sin{\lambda_{i} \xi}}{\xi} + \beta_{i}
    \frac{\cos{\lambda_{i} \xi}}{\xi} \right),
\end{equation}
with
\begin{equation}
    Y_{i}^2 = \frac{1}{2} \left( 1-\rho \right) \left( 1 + \beta_{i}^2 \right) - \frac{1}{4 
    \lambda_{i}} \left( \sin{2 \lambda_{i}} - \sin{2\lambda_{i} \rho} \right) 
    \left(  1 - \beta_{i}^2 \right) + \frac{\beta_{i}}{\lambda_{i}} \left( \sin^2{\lambda_{i}} -
    \sin^2{\lambda_{i} \rho} \right).
\end{equation}
Thus, the homogeneous solution $\psi_{h}$ is
\begin{equation}
    \psi_{h} = \sum_{i=1}^{\infty}h_{i} H_{i}(\xi)e^{-\lambda_{i}^2 s},
\end{equation}
with prefactors $h_{i}$ to be determined from the initial condition. 

In order to find a particular solution of the inhomogeneous problem, we first rewrite \ref{eqn_forpsi} as
\begin{equation}
    \frac{\partial \psi(\xi,s)}{\partial s} - \frac{1}{\xi^2} 
    \frac{\partial}{\partial \xi} \left( \xi^2 \frac{\partial \psi(\xi,s)}{\partial \xi} \right) =
    \mathcal{R}(\xi,s). \label{eqn_forpsii}
\end{equation}
Now, we express $\mathcal{R}(\xi,s)$, as 
well as the unknown inhomogeneous solution $\psi_{i}(\xi,s)$ in terms of the normalized eigenfunctions
$H(\xi,s)$ of the homogeneous problem,
\begin{equation}
    \mathcal{R}(\xi,s) = \sum_{i=1}^{\infty} R_{i}(s)H_{i}(\xi), \label{eqn_seriesforR}
\end{equation}
and
\begin{equation}
    \psi_{i}(\xi,s) = \sum_{i=1}^{\infty} C_{i}(s)H_{i}(\xi). \label{eqn_seriesforpsi}
\end{equation}
Substituting these forms into \ref{eqn_forpsii}, and noting that each term in the
series must vanish separately we obtain
\begin{equation}
    \frac{\partial C_{i}(s)}{\partial s} + \lambda_{i}^2C_{i}(s) - 
    R_{i}(s)  = 0.
\end{equation}
From the form of $\mathcal{R}(\xi,s)$ it follows that $R_{i}(s) = \alpha_{i}f_{0}e^{\sigma s}$ 
with some purely numerical prefactors $\alpha_{i}$, so we expect $C_{i}(s) \propto p_{i}e^{\sigma s}$
and find
\begin{equation}
    p_{i} = \frac{\alpha_{i} f_{0}}{\sigma + \lambda_{i}^2}.
\end{equation}
Finally, we determine the $\alpha_{i}$ by reconsidering Equation \ref{eqn_seriesforR}. We 
multiply both sides by $\xi^2 H_{j}(\xi)$, where $H_{j}(\xi)$ is one specific but arbitrary 
eigenfunction of the homogeneous problem, and then integrate over the whole volume $V$.
By the orthogonormality of these eigenfunctions we obtain
\begin{equation}
        \alpha_{j} = \int \frac{1}{1-\rho} \left( 3 - \frac{2\rho}{\xi} - 
        \sigma \left( \frac{1}{2} \xi^2 - \rho \xi + g_{0} \right) \right) \xi^2 H_{j}(\xi) d\xi,
\end{equation}
and all the $\alpha_{i}$ can be calculated explicitely. 
Thus, the full solution of the linear problem is
\begin{equation}
    c(\xi,s) = \sum_{i=1}^{\infty} \left( h_{i}e^{-\lambda_{i}^2 s} + 
    \frac{\alpha_{i}f_{0}}{\sigma + \lambda_{i}^2} e^{\sigma s} \right) H_{i}(\xi) + \frac{1}{1-\rho}
    \left( \frac{1}{2} \xi^2 - \rho \xi + g_{0} \right)f_{0} e^{\sigma s}.
\end{equation}
The constant $g_{0}$ can now be calculated from the requirement that $\int c(\xi,s=0) dV = N_{0}$. 
Here we make use of the fact that $\int H_{i}(\xi)\xi^2 d\xi = 0$ if $\lambda_{i}$ satisfies
Equation \ref{eqn_transcendental}, thus
\begin{equation}
    g_{0} = \frac{(1-\rho)/\sigma - \frac{1}{10} + \frac{1}{4}\rho 
    + \frac{1}{10}\rho^5 - \frac{1}{4} \rho^5 }{\frac{1}{3}\left(1-\rho^3\right) }.
\end{equation}







\end{document}